**Title:**

# Detecting Fake News using Machine Learning: A Systematic Literature Review

[1]**Alim Al Ayub Ahmed** (alim@jju.edu.cn)
[1]School of Accounting, Jiujiang University, Jiujiang, Jiangxi, CHINA

[2]**Ayman Aljarbouh**
[2]Department of Computer Science, University of Central Asia, 310 Lenin Street, 722918 Naryn, KYRGYZSTAN

[3]**Praveen Kumar Donepudi**
[3]Enterprise Architect, Information Technology, UST-Global, Inc., Ohio, USA

[4]**Myung Suh Choi**
[4]Monta Vista High School, 21840 McClellan Rd, Cupertino, CA 95014, USA

## Abstract

Internet is one of the important inventions and a large number of persons are its users. These persons use this for different purposes. There are different social media platforms that are accessible to these users. Any user can make a post or spread the news through these online platforms. These platforms do not verify the users or their posts. So some of the users try to spread fake news through these platforms. These fake news can be a propaganda against an individual, society, organization or political party. A human being is unable to detect all these fake news. So there is a need for machine learning classifiers that can detect these fake news automatically. Use of machine learning classifiers for detecting the fake news is described in this systematic literature review.

# 1. Introduction

World is changing rapidly. No doubt we have a number of advantages of this digital world but it also has its disadvantages as well. There are different issues in this digital world. One of them is fake news. Someone can easily spread a fake news. Fake news is spread to harm the reputation of a person or an organization. It can be a propaganda against someone that can be a political party or an organization. There are different online platforms where the person can spread the fake news. This includes the Facebook, Twitter etc. Machine learning is the part of artificial intelligence that helps in making the systems that can learn and perform different actions (Donepudi, 2019). A variety of machine learning algorithms are available that include the supervised, unsupervised, reinforcement machine learning algorithms. The algorithms first have to be trained with a data set called train data set. After the training, these algorithms can be used to perform different tasks. Machine learning is using in different sectors to perform different tasks. Most of the time machine learning algorithms are used for prediction purpose or to detect something that is hidden.

Online platforms are helpful for the users because they can easily access a news. But the problem is this gives the opportunity to the cyber criminals to spread a fake news through these platforms. This news can be proved harmful to a person or society. Readers read the news and start believing it without its verification. Detecting the fake news is a big challenge because it is not an easy task (Shu et al., 2017). If the fake news is not detected early then the people can spread it to others and all the people will start believing it. Individuals, organizations or political parties can be effected through the fake news. People opinions and their decisions are affected by the fake news in the US election of 2016 (Dewey, 2016).

Different researchers are working for the detection of fake news. The use of Machine learning is proving helpful in this regard. Researchers are using different algorithms to detect the false news. Researchers in (Wang, 2017) said that fake news detection is big challenge. They have used the machine learning for detecting fake news. Researchers of (Zhou et al., 2019) found that the fake news are increasing with the passage of time. That is why there is a need to detect fake news. The algorithms of machine learning are trained to fulfill this purpose. Machine learning algorithms will detect the fake news automatically once they have trained.

This literature review will answer the different research questions. The importance of machine learning to detect fake news will be proved in this literature review. It will also be discussed how machine learning can be used for detecting the false news. Machine learning algorithms that are used to detect false news will be discussed in the literature review.

The structure of the rest of paper is as Methodology in section two, section three shows the research questions, section four is showing the search process model that is followed for this literature review, result and discussion is given in section five, the conclusion is presented in section six. In the last, references are given for the papers that are discussed in this literature review.

# 2. Methodology

This literature review is written for answering some research questions. So the methodology that is used is the systematic literature review. This methodology helps in answering the research questions. The papers were collected from various databases to be discussed in this literature review. To answer the research questions, different research papers are discussed and cited in this literature review.

**Exclusion and Inclusion**

A number of papers are published every day. So when a string is searched a number of papers are presented in the result. Not all the papers are relevant to that string. This means there is a need for the criteria. The criteria for inclusion and exclusion that is followed in this literature review is given in the below table.

| Exclusion Criteria | Inclusion Criteria |
|---|---|
| The language of the paper is not the English language. | Papers that are written in the English language. |
| The complete paper is not accessible. | Paper can be accessed completely. |
| Paper is not related to machine learning and fake or false news detection. | Paper showing content related to machine learning and fake or false news detection. |

Table 1: Exclusion and Inclusion Criteria

Papers that fulfilled the above mentioned inclusion criteria were included in the literature review.

**Quality Assessment**

Quality of all included papers was assessed on the basis of the research work presented in those papers. The papers in which the researchers have discussed the machine learning use for fake or false news detection were considered as good quality papers to be included in this literature review.

# 3. Research Question

A SLR has to answer some RQs. In this literature review, three research questions will be answered on the basis of valid arguments. These two research questions are given below.

$RQ_1$: Why machine learning is required to detect the fake news?

$RQ_2$: Which machine learning supervised classifiers can be used for detecting fake news?

$RQ_3$: How classifiers of machine learning are trained to detect fake news?

These research questions will be answered in the result and discussion section of this literature review.

# 4. Search Process

A search process is followed to collect the papers that can be discussed in this literature review. This search process can easily be understood through the below given diagram.

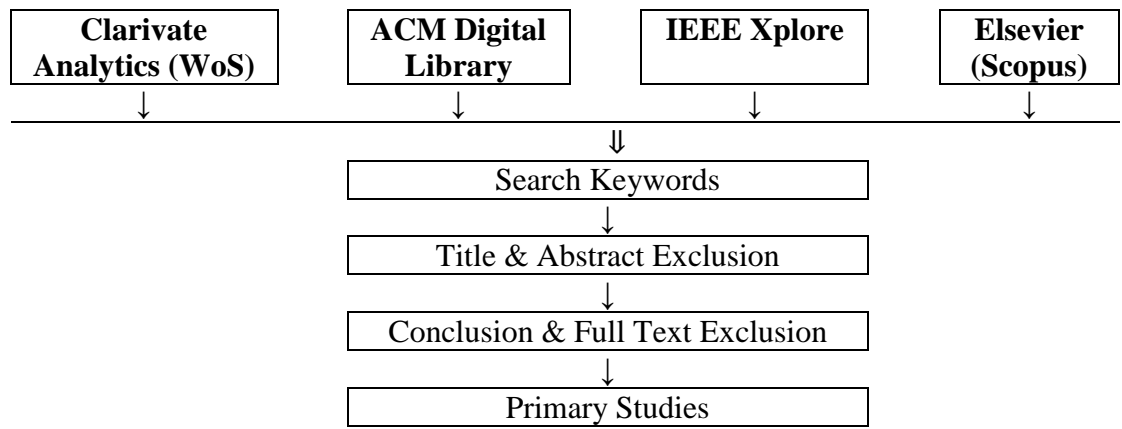

Figure 1: Search Process Model Diagram

Papers were collected from different databases. But not all of them were relevant to the topic. So first of all, the papers were excluded on the basis of their titles and abstract. An abstract is a kind of short summary of the whole paper that can give the idea about the contents presented in the paper. In the next phase, the further part of the papers was studied against the inclusion and exclusion criteria. Seventy three papers were collected from different databases against the search keyword. After the exclusion, there were twenty six papers were remaining that are discussed in this literature review.

## 5. Result and Discussion

Internet is one of the great sources of information for its users (Donepudi, 2020). There are different social media platforms that includes Facebook or Twitter that helps the people to connect with other people. Different kind of news are also shared on these platforms. People nowadays prefer to access the news from these platforms because these are easy to use and easy to access platforms. Another advantage to the people is that these platforms provide options of comments, reacts etc. These advantages attract people to use these platforms (Donepudi et al., 2020b). But as like their advantages, these platforms are also used as the best source by the cyber criminals. These persons can spread the fake news through these platforms. There is also a feature of sharing the post or news on these platforms and this feature also proves helpful for spreading such fake news. People start believing in such news as well as shares the news with other peoples. Researchers in (Zubiaga et al., 2018) said that it is difficult to control the false news from spreading on these social media platforms.

Anyone can be registered on these platforms and can start spreading news. A person can create a page as a source of news and can spread the fake news. These platforms do not verify the person whether he is really reputable publisher. In this way, anyone can spread news against a person or an organization. These fake news can also harm a society or a political party. The report shows that it is easy to change people opinions by spreading fake news (Levin, 2017). Therefore, there is a need for detecting these fake news from spreading so that the reputation of a person, political party or an organization can be saved.

**RQ$_1$: Why machine learning is required to detect the fake news?**

Increasing use of internet has made it easy to spread the false news. Different social media platforms can be used to spread fake news to a number of persons. With the share option of these platforms, the news spread in a fast way. Fake news just not only affects an individual but it can also affect an organization or business (Donepudi et al., 2020a). So controlling the fake news is mandatory. A person can know the news is fake only when he knows the complete story of that topic. It is a difficult task because most of the people do not know about the complete story and they just start believing in the fake news without any verification.

The question arises here how to control fake news because a person cannot control the fake news. The answer is machine learning. Machine learning can help in detecting the fake news (Khan et al., 2019). Through the use of machine learning these fake news can be detected easily and automatically (Della Vedova et al., 2018). Once someone will post the fake news, machine learning algorithms will check the contents of the post and will detect it as a fake news. Different researchers are trying to find the best machine learning classifier to detect the fake news (Kurasinski, 2020). Accuracy of the classifier must be considered because if it failed in detecting the fake news then it can be harmful to different persons. The accuracy of the classifier depends on the training of this classifier. A model that is trained in a good way can give more accuracy. There are different machine learning classifiers are available that can be used for detecting the fake news that will be answered in the next question.

**RQ 2: Which machine learning supervised classifiers can be used for detecting fake news?**

Detecting the fake news is one of the most difficult tasks for a human being. The fake news can easily be detected through the use of machine learning. There are different machine learning classifiers that can help in detecting the news is true or false. Nowadays, the dataset can easily be collected to train these classifiers. Different researchers used machine learning classifiers for checking the authenticity of news. Researchers in (Abdullah-All-Tanvir et al., 2019) used the machine learning classifiers for detecting the fake news. According to the experiments of the researchers the SVM and Naïve Bayes classifiers are best for detecting fake news. These two are better than other classifiers on the basis of accuracy they provide. A classifier with more accuracy is considered as a better classifier. The major thing is the accuracy that is provided by any classifier. Classifier with more accuracy will help in detecting more fake news. Researchers in (Kudarvalli & Fiaidhi, 2020) said that detection of false news is necessary because many persons spread the fake news of social media to mislead the people. To safe the individuals or organizations from losing their reputation because of false news it is necessary to detect it (Rahman et al., 2020). They have said that the machine learning is very helpful in this regard. They used the different machine-learning algorithms and they also found that the Logistic regression is a better classifier because it gives more accuracy.

Researchers in (Aphiwongsophon & Chongstitvatana, 2018) said that the social media produce a large number of posts. Anyone can register on these platforms can do any post. This post can contain false information against a person or business entity. Detecting such false news is an important and also a challenging task. For performing this task the researchers have used the three machine learning methods. These are the Naïve Bayes, Neural network and the SVM. The accuracy provided by the Naïve Bayes was 96.08%. On the other hand, the other two methods that are neural network and SVM provided the accuracy of 90.90%.

According to the researchers of (Ahmed et al., 2017), false news has major impact on the political situation of a society. False news on the social media platforms can change opinions of peoples. People change their point of view according to a fake news without verifying it. There is a need for a way that can detect such news. The researchers have used classifiers of machine learning for this purpose. The classifiers that are used by different researchers are the K-Nearest Neighbor, Support Vector Machine, Logistic Regression, Linear Support Vector Machine, Decision tree, Stochastic Gradient Descent. According to results, linear support vector machine provided the good accuracy in detecting the false news.

Researchers (Reis et al., 2019) have used the machine-learning classifiers for the detection of fake news. They have used different features to train these classifiers. Training of the classifiers is an important task because a trained classifier can give the more accurate results. According to the researchers of (Granik & Mesyura, 2017), artificial intelligence is better to detect the fake news. They have used Naïve Bayes classifier to detect fake news from Facebook posts. This classifier has given them the accuracy of 74% but they said the accuracy can be improved. To improve the accuracy different ways are also described by these researchers in that paper. There are classifiers of machine learning that are used for detecting fake news.

Some of these popular classifiers are given below that are used for this purpose.

**Support Vector Machine:** This algorithm is mostly used for classification. This is a supervised machine learning algorithm that learns from the labeled data set. Researchers in (Singh et al., 2017) used various classifiers of machine learning and the support vector machine have given them the best results in detecting the fake news.

**Naïve Bayes:** Naïve Bayes is also used for the classification tasks. This can be used to check whether the news is authentic or fake. Researchers in (Pratiwi et al., 2017) used this classifier of machine learning to detect the false news.

**Logistic Regression:** This classifier is used when the value to be predicted is categorical. For example, it can predict or give the result in true or false. Researchers in (Kaur et al., 2020) have used this classifier to detect the news whether it is true or fake.

**Random Forests:** In this classifier, there are different random forests that give a value and a value with more votes is the actual result of this classifier. In (Ni et al., 2020) researchers have used different machine learning classifiers to detect the fake news. One of these classifiers is the random forest.

**Recurrent Neural Network:** This classifier is also helpful for detecting the fake news. Researchers in (Jadhav & Thepade, 2019) have used the recurrent neural network to classify the news as true or false.

**Neural Network:** There are different algorithms of machine learning that are used to help in classification problems. One of these algorithms is the neural network. Researchers in (Kaliyar et al., 2020) have used the neural network to detect the fake news.

**K-Nearest Neighbor:** This is a supervised algorithm of machine learning that is used for solving the classification problems. This stores the data about all the cases to classify the new case on the base of similarity. Researchers (Kesarwani et al., 2020) have used this classifier to detect fake news on social media.

**Decision Tree:** This supervised algorithm of machine learning can help to detect the fake news. It breaks down the dataset into different smaller subsets. Researchers in (Kotteti et al., 2018) have used different machine learning classifiers and one of them is the decision tree. They have used these classifiers to detect the fake news.

**RQ3: How machine learning classifiers are trained for detecting fake news?**

Training of the classifiers of machine learning is an important task. This plays an important role for the accuracy of results of these classifiers. A classifier must have to be trained in a proper way with proper data set. Different researchers have trained the machine learning classifiers to detect the fake news. The main problem that occurs while training these classifiers is that mostly the training data set in an imbalanced form (Wang et al., 2020). Researchers in (Al Asaad & Erascu, 2018) have used the supervised machine learning classifiers for fake news detection. To train these classifiers they have used the three different models for feature extraction. Actually, these features are used to train the classifiers. These models are the TF-IDF Model, N-Gram Model, Bag of Words Model. These models extract the features from the training data set and then the classifier is trained through these features. Researchers in (Ahmed et al., 2018) has trained some machine learning classifiers to detect the fake news. For the training purpose, they have used a training data set. They have first removed the unnecessary words and the words are transformed to its single form. So that the training dataset that is given to these classifier should only have the valuable data.

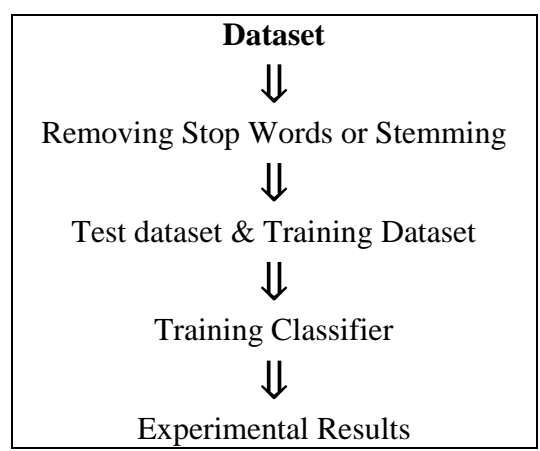

Figure 2: Training Dataset

Figure 3 shows the steps that are used while training a classifier. After the training a classifier is then used for experiments.

## 6. Conclusion

Due to increasing use of internet, it is now easy to spread fake news. A huge number of persons are regularly connected with internet and social media platforms. There is no any restriction while posting any news on these platforms. So some of the people takes the advantage of these platforms and start spreading fake news against the individuals or organizations. This can destroy the repute of an individual or can affect a business. Through fake news, the opinions of the people can also be changed for a political party. There is a need for a way to detect these fake news. Machine learning classifiers are using for different purposes and these can also be used for detecting the fake news. The classifiers are first trained with a data set called training data set. After that, these classifiers can automatically detect fake news.

In this systematic literature review, the supervised machine learning classifiers are discussed that requires the labeled data for training. Labeled data is not easily available that can be used for training the classifiers for detecting the fake news. In future a research can be on the use of the unsupervised machine learning classifiers for the detection of fake news.

--0--